\documentstyle[12pt,A4]{article}

\begin{document}
\makeatletter
\def\siml{\mathrel{\mathpalette\gl@align<}}
\def\simg{\mathrel{\mathpalette\gl@align>}}
\def\gl@align#1#2{\lower.6ex\vbox{\baselineskip\z@skip\lineskip\z@
 \ialign{$\m@th#1\hfill##\hfil$\crcr#2\crcr{\sim}\crcr}}}
\makeatother
\def\thefootnote{\fnsymbol{footnote}}
\hbadness=10000
\hbadness=10000
\begin{titlepage}
\nopagebreak
\begin{flushright}

        {\normalsize
 Kanazawa-94-09\\
 KITP-94-02\\

May, 1994   }\\
\end{flushright}
\vspace{.2cm}
\vfill
\begin{center}
\renewcommand{\thefootnote}{\fnsymbol{footnote}}
{\large \bf Minimal String Unification and Constraint on Hidden Sector}

\vfill
\vspace{0.4cm}

{\bf Hiroshi Kawabe,
Tatsuo Kobayashi$^*$
\footnote[2]{e-mail:kobayasi@hep.s.kanazawa-u.ac.jp}
 } and
{\bf Noriyasu Ohtsubo$^{**}$
\footnote[3]{e-mail:ohtsubo@neptune.cisp.kanazawa-it.ac.jp}
}\\

\vspace{0.4cm}
Yonago National College of Technology, \\
Yonago, 683, Japan \\

      $^*$Department of Physics, Kanazawa University, \\
       Kanazawa, 920-11, Japan \\

$^{**}$Kanazawa Institute of Technology, \\
Ishikawa 921, Japan

\vfill
\end{center}
\vspace{0.4cm}

\vfill
\nopagebreak
\begin{abstract}
We examine whether a minimal string model possessing the same massless spectra
 as the MSSM can be obtained from $Z_4$, $Z_6$ and $Z_8$ orbifold
 constructions.
Using an anomaly cancellation condition of the target space duality symmetry,
 we derive allowable values of a level $k_1$ of U(1)$_Y$ for the minimal
 string model on the orbifolds through computer analyses.
We investigate threshold corrections of the gauge coupling constants of SU(3),
 SU(2) and U(1)$_Y$ and examine consistencies of the model with the LEP
 experiments.
It is found that $Z_4$ and $Z_8$-II can not derive the minimal string model
 but $Z_6$-I, $Z_6$-II and $Z_8$-I are possible to derive it with $k_1=29/21$,
 $1\leq k_1\leq 32/21$ and $1\leq k_1\leq 41/21$ respectively.
We obtain explicitly allowed combinations of modular weights.
The minimum values of the moduli on unrotated planes are estimated within the
 ranges of the levels.
Further we investigate what kinds of hidden sectors are consistent with the
minimal string models.
Also their gauge coupling constants of the hidden groups are estimated.
We discuss Yukawa couplings of the models.

\end{abstract}

\vfill
\end{titlepage}
\pagestyle{plain}
\newpage
\voffset = 0.5 cm

\vspace{0.8 cm}
\leftline{\large \bf 1. Introduction}
\vspace{0.8 cm}

Superstring theories are the only known candidates for unified theories of
all the interactions including gravity.
All the gauge coupling constants are unified even without a unified group at a
string scale $M_{\rm st}=5.27\times g_{\rm st} \times 10^{17}$GeV
\cite{Kaplunovsky}, where $g_{\rm st}\simeq 1/\sqrt 2$ is a universal string
coupling constant.
There are crucial problems how to derive the minimal supersymmetric standard
model (MSSM) at the low energy from superstring theories and how to break the
supersymmetry (SUSY).
For the first problem, several types of scenarios have been discussed to lead
to the MSSM spectrum.
Some of them have intermediate scales of grand unified theories (GUTs)
like SU(5) or SO(10), and others have extra matter fields other than the MSSM
spectrum.
However GUTs often face problems on triplet-doublet splitting in the Higgs
sector in addition to mass spliting of the quarks and
leptons other than the third generation.
Further some models with extra matter fields lead to the fast proton decay.
Here we concentrate ourselves to  the simplest scenario that the massless
spectrum at $M_{\rm st}$ is same as one of the MSSM.
That does not suffer the above problems and is called a minimal string model.
Explicit searches for the model have been done, e.g., within the framework
of orbifold models.

The construction of the 4-dim string vacua through the orbifolds is one of
the simplest methods among several types of constructions \cite{ZNOrbi}.
For the orbifold models, we can obtain several phenomenological aspects
such as Yukawa couplings, K\"ahler potentials, threshold corrections of the
gauge coupling constants and so on.
Although some standard-like models have been found in the orbifold models
\cite{ZNOrbi2}, a completely realistic model has never been derived.
It seems that needed are phenomenological constraints to obtain the realistic
model from a huge number of 4-dim string vacua.

Recent study of the LEP measurements shows that all the gauge couplings of the
MSSM are unified at $M_X \sim 10^{16}$GeV \cite{MSSM}.
A difference between $M_X$ and $M_{\rm st}$ seems to reject the possibility
for the minimal string model.
However it is expected that this difference is explained by threshold
corrections of the gauge couplings due to higher massive modes of the string.
The corrections have been calculated in the orbifold models
\cite{Dixon,Derendinger}.
The corrections as well as masses depend on moduli $T$, whose vacuum
expectation values describe orbifold geometries.
Large values of the moduli lead to large threshold corrections, although
a fairly larger moduli value seems unnatural.
Refs.\cite{ILR,Ibanez} studied the threshold corrections consistent with the
measured values of the gauge couplings at $M_Z$ in the case where a Kac-Moody
level $k_1$ of $U(1)_Y$ is equal to 5/3.
This value of $k_1$ is predicted by GUTs, but string theories could derive any
other level \cite{Ibanez2}.
Thus, it is very interestring to extend to the cases of general levels.
In ref.\cite{KKO2}, the levels of $Z_N\times Z_M$ orbifold models are
estimated as $1\leq k_1 \siml 2$ in order to
explain the measured values of the gauge coupling constants at $M_Z$.
In this paper we study in detail the possibility for obtaining the minimal
string model from $Z_N$ orbifolds in the case where $1\leq k_1 <2$.

In the above discussion a target-space duality symmetry \cite{Kikkawa} is
very important and that is a \lq \lq stringy" feature.
Also an effective lagrangian is invariant under the symmetry of the moduli
fields.
However loop effects due to only massless modes make the symmetry anomalous
\cite{Derendinger}.
The duality anomaly can be cancelled by two ways.
One is the Green-Schwarz mechanism \cite{GS,Derendinger2,Derendinger},
which induces a nontrivial
transformation of a dilaton field $S$ under the duality transformation.
Further the anomaly can be cancelled in terms of the moduli dependent
threshold corrections of the gauge couplings due to the massive modes.
The former is independent of gauge groups and it constrains strongly the
massless spectrum, such as gauge anomalies constrains spectra in field
theories.\footnote{
Note that the $Z_N \times Z_M$ orbifold models do not have such a constraint,
because all the moduli of the three planes contribute the threshold
corrections.}
The strong constraint makes possible to investigate massless spectra
instead of the explicit search by shifts and Wilson lines \cite{WL1,WL2,KO} on
an E$_8 \times $E$_8'$ lattice.
The anomaly cancellation condition forbids the minimal string model in
$Z_3$ and $Z_7$ orbifild models \cite{Ibanez}.
We do not study here the $Z_3$ and $Z_7$ orbifold models.
Further we omit the $Z_{12}$ orbifold models, which have several types of
twisted sectors.
Thus the $Z_{12}$ orbifold models need longer analysis.
However we can study the $Z_{12}$ orbifold models in a way similar to the
following discussion.

Hidden sectors play a role in the SUSY-breaking.
A gaugino condensation in the hidden sector is one of the realistic
SUSY-breaking mechanisms \cite{cond1}, which
determines vacuum expectation values of the dilaton and moduli fields
\cite{cond2}.
The expectation value of $S$ derives a gauge coupling constant.
Refs.\cite{cond2} show that a scalar potential of the moduli fields has a
minimum around a self-dual point of the duality.
Also effective one-loop potential is considered in refs.\cite{Ross}.
The gaugino condensation around $10^{13}$GeV could lead to soft
SUSY-breaking terms at the weak scale.
The duality anomaly cancellation condition restricts the hidden sector so as
to be consistent with the MSSM as discussed in ref.\cite{koba}, where
the duality anomaly cancellation condition on U(1)$_Y$ was not taken into
account.
In this paper we study the allowed gauge groups and matter fields in hidden
sectors of the minimal string models derived from the $Z_6$-I, $Z_6$-II and
$Z_8$-II orbifolds.
The consistency with U(1)$_Y$ is investigated in more detail.
Also estimated are their gauge coupling constants at $10^{13}$GeV and their
blow-up scales.

This paper is organized as follows.
In section two we review on the orbifold models.
Also their massless conditions are studied to find conditions on
oscillation numbers of the MSSM matter fields under some values of the
level $k_1$.
In section three we review on the duality symmetry and the threshold
corrections.
The discussion of section two leads to allowed modular weights for each
MSSM matter field.
The duality anomaly cancellation condition is also reviewed.
In section four we investigate the possibility to derive the minimal
string model with consistent values of the measured gauge couplings from
the $Z_N$ orbifold models.
We try to assign the allowed modular weights to matter fields of the MSSM
and find combinations of the modular weights to satisfy the duality anomaly
cancellation condition for the SU(3), SU(2) and U(1)$_Y$ gauge groups.
Through such an analysis we get allowed values of the level $k_1$.
Then we study whether the allowed combinations lead to the threshold
corrections consistent with the measurements.
Some tables show explicitly the combinations, which are useful for model
building.
Here we restrict ourselves to the case where the vacuum expectation values of
the muduli fields are of order one.
In section five we find the hidden sectors which are consistent with
the observable massless spectra obtained in section four from viewpoint of
the duality anomaly cancellation conditions.
That does not allow hidden sectors which have smaller number of matter
fields with non-trivial representations under hidden gauge groups.
Then we estimate their hidden gauge coupling constants at $10^{13}$GeV and
their blow-up scales.
In section six we study Yukawa couplings of the allowed models, explicitly.
That could also be used for phenomenological constraints.
Section seven is devoted to conclusions and discussions.

\vspace{0.8 cm}
\leftline{\large \bf 2. $Z_N$ Orbifold Models}
\vspace{0.8 cm}

In the orbifold models \cite{ZNOrbi}, the string states consist of the bosonic
strings on the 4-dim space-time and a 6-dim orbifold, their right-moving
superpartners and left-moving gauge parts whose momenta span a shifted
E$_8 \times $E$_8'$ lattice.
The right-moving fermionic parts are bosonized and momenta of the bosonized
 fields span an SO(10) lattice.
The 6-dim orbifolds are obtained through the division of a 6-dim space R$^6$
 by 6-dim Lie lattices and their automorphisms (twists).
We use here an SO(5)$^2\times $SU(2)$^2$ lattice
for the $Z_4$ orbifold, a G$^2_2 \times$ SU(3) lattice for the $Z_6$-I,
a G$_2 \times $SU(3)$\times$SU(2)$^2$ lattice for the $Z_6$-II,
an SO(9)$\times$SO(5) lattice for the $Z_8$-I and
an SO(9)$\times$SU(2)$^2$ lattice for the $Z_8$-II \cite{KO}.
Using other lattice, we can construct these orbifolds.
For any lattice, we can study in a way similar to the following discussion.
We denote eigenvalues of the twist $\theta$ in a complex basis
 ($X_i,\tilde X_i$) ($i=1,2,3$) as exp$[2\pi i v^i]$, whose exponents $v^i$
are
(1,1,2)/4 for the $Z_4$ orbifold, (1,1,4)/6 for the $Z_6$-I, (1,2,3)/6 for
$Z_6$-II, (1,5,2)/8 for the $Z_8$-I and (1,3,4)/8 for the $Z_8$-II.
The twist $\theta$ is embedded into the SO(10) and
 E$_8 \times $E$_8'$ lattices in terms of shifts so that the $N=1$ SUSY
remains and the gauge group breaks into a small one.
The E$_8 \times $E$_8'$ lattice is shifted by Wilson lines, as well.

There are two types of closed strings on the orbifolds.
One is an untwisted string whose massless states should satisfy
$$h-1=0, \eqno(2.1)$$
where $h$ is a conformal dimension of the E$_8\times$E$'_8$ gauge part.
The other is a twisted string.
Massless states of $\theta^\ell$-twisted sector should satisfy
 the following condition:
$$h+N_i+c_{\ell}-1=0, \eqno(2.2)$$
where $N_i$ is an oscillation number associated with the $i$-th plane
and $c_{\ell}$ is obtained from
$$ c_{\ell}= {1\over 2}\sum_{i=1}^3 v^i_{\ell}(1-v^i_{\ell}),
\qquad v^i_{\ell} \equiv \ell v^i-{\rm Int}(\ell v^i).
\eqno(2.3)$$
Here ${\rm Int}(a)$ represents an integer part of $a$.

A representation $\underline{R}$ of the non-abelian group $G$ contributes to
 the conformal dimension as
$$ h={C(\underline{R}) \over C(G)+k},\eqno(2.4)$$
where $k$ is a level of a Kac-Moody algebra and
 $C(G)$ ($C(\underline{R})$) is a quadratic Casimir of the adjoint
 ($\underline{R}$) representation of the group $G$, e.g., $C({\rm SU}(N))=N$
($C(\underline{N})=(N^2-1)/2N$), $C({\rm SO}(2n))=2n-2$
($C(\underline{2n})=(2n-1)/2$),  $C($E$_6)=12$ ($C(\underline{27})=26/3$),
 $C($E$_7)=18$ ($C(\underline{56})=57/4$) and $C($E$_8)=30$.
In general the string theories derive the gauge groups with $k=1$,
\footnote{Gauge groups with $k\not=1$ are discussed in refs.~\cite{k}.}
 except for U(1).
Then we restrict ourselves to the case where $k=1$ for the non-abelian groups.
It follows that the representations $\underline{N}$ of SU($N$),
 $\underline{2n}$ (vector) of SO($2n$), $\underline{27}$ of E$_6$ and
 $\underline{56}$ of E$_7$ group have the conformal dimensions of
 $h=(N-1)/2N$, 1/2, 2/3 and 3/4, respectively.
A state with a charge $Q$ of the U(1)$_Y$ has an additional contribution
 of $h=Q^2/k_1$ where $k_1$ is the level of U(1)$_Y$.

It is found that all the above representations are possible to satisfy the
 massless condition (2.1) in the untwisted sector if they have suitable
 charges of the extra U(1)'s.
However, they are not always satisfy the massless condition (2.2) in the
twisted sector.
We can get higher bounds of the oscillation number $N_i$ for each
representations from the condition (2.2).
For the MSSM matter fields, the bounds depend on the level $k_1$.
For example we consider a quark singlet with U(1)$_Y$ charge $Q$.
For this state to have $N_i$, the level $k_1$ should satisfy the
following bound:
$$k_1\geq {Q^2 \over 2/3-c_\ell-N_i}.
\eqno(2.5)$$
For the other MSSM matter fields, we have similar relations.
Existence of each MSSM matter field in the untwisted sector gives a lower
bound of $k_1$.
We have $k_1\geq 1$ so that singlets with $Q=1$ appear.

\vspace{0.8 cm}
\leftline{\large \bf 3. Duality and Threshold Corrections}
\vspace{0.8 cm}

The duality symmetry is retained in effective field
 theories derived from the orbifold models \cite{Ferrara}.
In the theories, moduli fields $T_i$ ($i=1,2,3$) associated with the $i$-th
 complex planes have the K\"ahler potentials
$$ -\sum_i{\rm log}|T_i+\bar T_i|,\eqno(3.1)$$
which are invariant under a duality transformation:
$$ T_i \rightarrow {a_iT_i-ib_i \over ic_iT_i+d_i} ,\eqno(3.2)$$
up to the K\"ahler transformation, where $a_i,b_i,c_i,d_i\in {\bf Z}$ and
 $a_id_i-b_ic_i=1$.

The K\"ahler potential of the matter field $A$ is
$$ \prod^3_{i=1}(T_i+\bar T_i)^{n^i}A\bar A,\eqno(3.3)$$
whose duality invariance requires the following transformation:
$$ A \rightarrow A \prod_{i=1}^3(ic_iT_i+d_i)^{n^i},\eqno(3.4)$$
where $n^i$ is called a modular weight \cite{Dixon2,Ibanez}.

For the untwisted sector associated with the $p$-th plane, the matter fields
 have $n^i=-\delta^i_p$.
The $\theta^\ell$-twisted state without oscillators has the following
 modular weights:
$$\begin{array}{llllll}
n^i&=&v^i_{\ell}-1, & \qquad v^i_{\ell}& \neq & 0, \\
n^i&=&0,            & \qquad v^i_{\ell}& =    & 0.
\end{array}
\eqno(3.5)$$
The oscillator $\partial X_i$ reduces the corresponding element of the modular
 weight by one and the oscillator $\partial \tilde X_i$ contributes oppositely.
Thus we can obtain modular weights of the matter fields using the allowable
 values for $N_i$.

The allowed modular weights for the MSSM matter fields depend on the value of
the level $k_1$.
Table 1 lists the modular weights and the lower bounds of $k_1$ for each
 MSSM matter field permitted by the massless condition in the previous section
 for $\theta$- and $\theta^2$-twisted sectors in the $Z_4$ orbifold model.
(In the table $\theta^3$-twisted sector is omitted because it includes only
 anti-matters \cite{KO2}.
Such sectors are also omitted in the following tables.)
In the table, the underline represents any permutation of the elements.
For example, the representations $(3,2)_{1/6}$, $(\bar 3,1)_{1/3}$,
 $(\bar 3,1)_{-2/3}$, $(1,2)_{\pm 1/2}$ and $(1,1)_1$ in $\theta$ twisted
 sector are able to possess $n^i=(-3,-3,-2)/4$, if $k_1\geq 4/15$, $16/51$,
 $64/51$, $4/7$ and $16/11$ respectively.
The modular weight $n^i=(\underline{-7,-3},-2)/4$ is not realized for
 $(3,2)_{1/6}$, $(\bar 3,1)_{-2/3}$ and $(1,1)_1$ in the case where
$k_1 <2$.
It seems that the orbifold models compatible with the experiments derive
 $k_1 < 2$ as suggested in the previous study about the $Z_N\times Z_M$
 orbifolds \cite{KKO2}.
Then we limit the subsequent studies to the cases of $k_1<2$.
For the twisted sectors of $Z_6$-I, $Z_6$-II, $Z_8$-I and $Z_8$-II, the
 modular weights and the lower bounds of $k_1$  for the MSSM matter fields
are listed in Table 2, Table 3, Table 4 and Table 5, respectively, where we
omit the modular weights requiring $k_1 \geq 2$.
The $Z_{12}$ orbifold models have much more modular weights than the others.
That makes analysis on the $Z_{12}$ orbifolds lengthy.
However, we can obtain modular weights of the $Z_{12}$ orbifolds, similarly.

When we consider the hidden sectors, we do not have to take care of the level
$k_1$ of U(1)$_Y$.
The last column of Table 2 shows which of modular weights can be possessed by
the hidden matter fields with the $N$-dim representation of the $Z_6$-I
orbifold models.
Plus signs in the column denote that their representations
are permitted by the massless condition (2.2) for all SU($N)'$.
Bounds of $N$ are given for the representation $\underline{N}$ of SU($N)'$.
We remark that the $Z_8$ orbifold models are possible to have
$N\leq 8$ in SU($N$), while the $Z_6$ models are $N\leq 9$ in SU($N$)
by the explicit search of gauge groups in terms of shifts and Wilson lines
 \cite{Z46,Gauge}.
In addition to the modular weights in Table 2, the matter fields with
the $N$-dim representation of SU($N)'$ ($N\leq 3$) are allowed to have
$n^i=(\underline {-10,-4,-4})/6$, and doublets of SU($2)'$ are able to
possess the following modular weights,
$$(\underline{-23,-5},-2)/6, \quad (\underline{-17,-11},-2)/6, \quad
(\underline{-11,-5},4)/6, $$
$$(\underline{-9,-3},0)/6, \quad  (\underline{3,-3},0)/6.
$$

Similarly the last column of Table 3 shows the allowed hidden matter
fields with the modular weights.
In addition, the matter fields with the $N$-dim representation of SU($N)'$
($N\leq 3$) are possible to have $n^i=(-17,-4,-3)/6$ and $(-5,-10,-3)/6$, and
further doublets of SU($2)'$ are able to possess
$n^i=(\underline {-9},0,\underline{-3})/6$  and
$(\underline 3,0,\underline {-3})/6$.
For the $Z_8$-I orbifold models, all matter fields with the $N$-dim
representations of SU($N)'$ are allowed to have every modular weight shown in
Table 4.
In addition, the matter fields with the $N$-dim representation of SU($N)'$
($N\leq 3$) are possible to have the following modular weights,
$$(\underline{-31,-3},-6)/8, \quad (\underline{-15,-3},-14)/8,
\quad (\underline{-7,5},-6)/8.$$
Furthermore doublets of SU(2)$'$ are able to possess
$n^i=(\underline{-12,-4},0)/8$ and $(\underline{4,-4},0)/8$.

The duality symmetry becomes anomalous by loop effects of only massless modes
 \cite{Derendinger}.
Triangle diagrams contributing to the duality anomaly have two gauge bosons
and moduli dependent connections like K\"ahler or $\sigma$-model connections
as external lines, and massless fermions in addition to gauginos as internal
lines.
Duality anomaly coefficients $b'^i$ of a group $G$ are obtained from
$$b'^i=-C(G)+\sum_{\underline R} T(\underline R)(1+2n^i_{\underline R}),
\eqno(3.6)$$
where $T(\underline R)$ is an index given by
 $T(\underline R)=C(\underline R){\rm dim}(\underline R)/{\rm dim}(G)$, e.g.,
 $T(\underline R)=1/2$ for the $N$-dim fundamental representation of SU($N$).

The duality anomaly can be cancelled by two ways.
One is the Green-Schwarz (GS) mechanism \cite{GS}, which induces a non-trivial
transformation to the dilaton field $S$ under the duality as
$$S \rightarrow S -{1 \over 8 \pi^2}\sum^3_{i=1} \delta^i_{\rm GS}{\rm log}
(ic_iT_i+d_i),
\eqno(3.7)$$
where $\delta^i_{\rm GS}$ is a GS coefficient.
Note that the above mechanism is independent of gauge groups.
Further the duality anomaly can be cancelled in terms of moduli dependent
threshold corrections due to massive modes.
The corrections depend only on the moduli whose planes are unrotated under
some twist, because only $N=2$ sectors contribute to the moduli dependent
threshold corrections of the gauge coupling constants.
For the other planes, the duality anomaly should be cancelled only by the GS
mechanism.
It works, if the following condition is satisfied:
$$b'^j_3=b'^j_2=b'^j_1/k_1=b'^j_{\rm hid}, \eqno(3.8)$$
for $j$-th planes rotated under any twist,
where $b'^j_3$, $b'^j_2$ and $b'^j_1$ are the anomaly coefficients of SU(3),
SU(2) and U(1)$_Y$, respectively, for the minimal string model in the observed
sector, as discussed in ref.\cite{Ibanez}.
The anomaly coefficient in the hidden sector is represented by
$b'^j_{\rm hid}$.
Only the first plane is concerned with eq.~(3.8) in the $Z_6$-II
orbifold models, whereas the first and the second planes are in $Z_4$,
$Z_6$-I, $Z_8$-I and $Z_8$-II orbifold models.
The equation (3.8) gives more stringent constraints to the latter orbifold
 models because $k_1$ is common for the both planes.

The other $k$-th ($k\not= j$) planes contribute in the threshold corrections
 of the gauge coupling constants induced by the tower of higher massive modes.
The threshold corrections \cite{Dixon,Derendinger} are given by
$$\Delta_a(T_k)=-{1 \over 16\pi^2}\sum_k (b'^k_a-k_a\delta^k_{\rm GS})
{\rm log}|\eta(T_k)|^4, \eqno(3.9)$$
where $\eta(T)=e^{-\pi T/12}\prod_{n\ge 1}(1-e^{-2\pi nT})$
 is the Dedekind function.
Eq.~(3.9) may be modified in the cases of the orbifolds
constructed through other lattices said above \cite{Mayr}.
Also presence of non-vanishing Wilson lines modify the threshold corrections
as well as the duality symmetry \cite{WL3}.
However the Wilson lines vanish for the $N=2$ sector \cite{WL2,KO}.
Thus we do not need the modification.
The moduli $T_2$ and $T_3$ participate in eq.~(3.9) for the $Z_6$-II orbifold,
 while only $T_3$ for the other orbifolds.
For simplicity, we consider a case of $T_2 = T_3(=T_k)$ in the $Z_6$-II
orbifold models.
Then we denote $\delta^k_{\rm GS}=\delta^2_{\rm GS}+\delta^3_{\rm GS}$,
 $b'^k_a=b'^3_a+b'^2_a$ and $n^k_{\underline R}=n^2_{\underline R}
+n^3_{\underline R}$ for the $Z_6$-II and remove the summation.

Using the threshold corrections, we obtain the one-loop coupling constants
 $\alpha_a(\mu)=k_ag_a^2(\mu)/4\pi$ ($k_3=k_2=k_{\rm hid}=1$) at an energy
scale $\mu$ as follows,
$$ \alpha^{-1}_a(\mu)=\alpha^{-1}_{\rm st}+{1 \over 4\pi}{b_a\over k_a}
 {\rm log}{M_{\rm st}^2 \over \mu^2}-{1\over 4\pi}
 ({b'^k_a \over k_a}-\delta^k_{\rm GS}){\rm log}[(T_k+\bar T_k)|\eta(T_k)|^4],
 \eqno(3.10)$$
where $\alpha_{\rm st}=g^2_{\rm st}/4\pi$ and $b_a$ are $N=1$
 $\beta$-function coefficients.
We use the same $b_3=-3$, $b_2=1$ and $b_1=11$ as ones of the MSSM.
There are also moduli-independent threshold corrections \cite{Kaplunovsky}.
It is expected that moduli-independent corrections are smaller than
moduli-dependent one.
In this paper we neglect moduli-independent corrections, although
the neglect leads to small uncertainty.

Here we discuss the unification of SU(3) and SU(2) gauge
couplings.
The renormalization group equation (3.10) relates the unification scale $M_X$
of SU(3) and SU(2) with $M_{\rm st}$ as follows \cite{Ibanez},
$$ {\rm log}{M_{X} \over M_{\rm st}}={1\over 8} \Delta b'^k
 {\rm log}[(T_k+\bar T_k)|\eta(T_k)|^4] ,
\eqno(3.11)$$
where $\Delta b'^k\equiv b'^k_3-b'^k_2$.
Note that the U(1)$_Y$ gauge coupling does not necessarily
 unify with SU(3) and SU(2) at $M_X$, because $k_1$ is not always equal to
$k_1=5/3$.
The term ${\rm log}[(T_k+\bar T_k)|\eta(T_k)|^4]$ is negative for any value of
$T$.
Thus to obtain $M_X<M_{\rm st}$, the anomaly coefficients should satisfy
$\Delta b'^k>0$.
Assuming the SUSY is broken at $M_Z$, we have
$$\log {M_X^2 \over M_Z^2}=\pi \left( \sin^2 \theta_W(M_Z)
\alpha^{-1}_{\rm em}(M_Z)-\alpha_3^{-1}(M_Z)\right).
\eqno(3.12)$$
Then we use
$\sin^2\theta_W(M_Z)=0.2325\pm .0008$,
$\alpha^{-1}_{\rm em}(M_Z)=127.9\pm .1$,
$\alpha_3^{-1}(M_Z)=8.82\pm .27$ at $M_Z=91.173\pm .020$ to obtain
$$M_X=10^{16.23\pm .27}\ {\rm GeV}, \qquad \alpha^{-1}_X=24.51\pm .09,
\eqno(3.13) $$
where $\alpha_X \equiv \alpha_3(M_X)=\alpha_2(M_X)$.
In the case where $\Delta b'^k=3$, we obtain $T\simeq 11$ through (3.11) using
$M_X=10^{16.2}$GeV and $M_{\rm st}=10^{17.6}$GeV.

We eliminate $\alpha_{\rm st}$, $\delta^k_{\rm GS}$ and $T_k$ in (3.10) for
$\alpha^{-1}_3$, $\alpha^{-1}_2$ and $\alpha^{-1}_1$ to obtain
\renewcommand{\arraystretch}{1.5}
\arraycolsep=0.5mm
$$ \begin{array}{rll}
(k_1b'^k_2-b'^k_1)\alpha_3^{-1}(\mu)&=& (b'^k_2-b'^k_3)\alpha_{\rm em}^{-1}
(\mu)-\left\{ b'^k_1+b'^k_2-(k_1+1)b'^k_3\right\} \alpha_{\rm em}^{-1}(\mu)
\sin^2 \theta_{\rm W}(\mu)\\
&+&{\displaystyle {1\over 4\pi}\left\{ 4b'^k_1-(3k_1+11)b'^k_2-(k_1-11)b'^k_3
\right\} {\rm log}{M_{\rm st}^2\over \mu^2}}.
\end{array} \eqno(3.14)$$

\vspace{0.8 cm}
\leftline{\large \bf 4. Minimal String Model}
\vspace{0.8 cm}

In this section we study the possibility for the minimal string model, using
the $Z_N$ orbifold models.
For the purpose, we use the duality anomaly cancellation condition instead of
searching explicitly massless spectra in terms of all the possible shifts and
the Wilson lines on the E$_8 \times $E$_8'$ lattice.
First of all, we assign allowed modular weights to the MSSM matter fields,
i.e.,
the three $(3,2)_{1/6}$ representations, three $(\bar 3,1)_{-2/3}$, three
$(\bar 3,1)_{1/3}$, five $(1,2)_{\pm 1/2}$ and three $(1,1)_{1}$, to find
combinations of the modular weights satisfying the duality anomaly
cancellation condition (3.8).
Note that the presence or absence of the right-handed neutrinos does not affect
 the following discussion, because they are singlets with $Q=0$ under the
standard gauge group.
As shown in ref.\cite{Ibanez}, we cannot derive any combination satisfying
(3.8) from the $Z_3$ and $Z_7$ orbifold models, whose orbifolds are
constructed through SU(3)$^3$ and SU(7) lattices and have exponents
$v^i=(1,1,1)/3$ and $(1,2,4)/7$, respectively.

For each allowed combination, we obtain the level $k_1$ by a ratio of the
anomaly coefficients, i.e., $k_1=b'^j_1/b'^j_3$.
Here we have to check whether or not each combination includes modular weights
allowed by this level $k_1$.

Next we study the possibility that the threshold corrections due to massive
modes derive the measured gauge coupling constants.
Namely we investigate whether or not combinations of the modular weights
allowed at the above stage derive
the gauge coupling constants consistent with all measurements falling
within the error bars, through (3.14).
Here we estimate $\alpha_3^{-1}(M_Z)$ through (3.14) using the value of
$k_1$ obtained through (3.8) and the measured values of
$\sin^2\theta_W(M_Z)$ and $\alpha_{\rm em}(M_Z)$, then compare it with the
experimental value $\alpha_3^{-1}(M_Z)$ in order to investigate the
consistency of the model.\footnote{Using (3.14) and experimental values of
the gauge couplings $\alpha_a^{-1}(M_z)$ ($a=1,2,3$), we can estimate the
value of $k_1$ through (3.14) \cite{KKO2}.
The value $k_1$ can be compared with $k_1$ obtained by (3.8) to investigate
the consistency of the models.
This analysis derives the same results as the procedure studied here.}
Here we restrict ourselves to the case where $T_k\leq 11$ and
$\Delta b'^k \geq 3$, in order to investigate the possibility that $T_k$ is
of order one.
However, we can analyze other case in a similar way.
The $Z_4$ and $Z_8$-II orbifold models cannot lead to the
minimal string model consistent with the measured gauge coupling constants.
For example $Z_4$ orbifold models lead $\alpha^{-1}_3(M_Z) \leq 4.0$ through
(3.14) using the measured values of sin$^2\theta_W(M_Z)$ and
$\alpha_{\rm em}(M_Z)$.
For the $Z_6$-I orbifold models, we have only one combination of modular
weights consistent with the measurements in the case where
$\Delta b'^k\geq 3$.
Even for the case $0 \leq \Delta b'^k\leq 3$, we can find only five
combinations, which are shown in Table 6.
In the table the MSSM matter fields are represented by $Q$, $U$, $D$, $L$,
$E$ and $H$ for the quark doublets, the quark singlets of the up-sector and
 down-sector, the lepton doublets, lepton singlets and the Higgs fields,
respectively.
In the table modular weights for the matter fields are represented by the
following numbers:
$$ 1:(-1,0,0), \quad 2:(0,-1,0), \quad 3:(0,0,-1),
\eqno(4.1)$$
$$ 4:(-5,-5,-2)/6, \quad 5:(-3,-3,0)/6, \quad 6:(-4,-4,-4)/6,$$
$$ 7:(-11,-5,-2)/6, \quad 8:(-5,-11,-2)/6, \quad 9:(-5,-5,4)/6.$$

For the $Z_8$-I orbifold models, 104 combinations with $\Delta b'^k\geq 3$ are
allowed, including 52 combinations with $\Delta b'^k=4$, which is the largest
value among them and leads to $T_k\simeq 8.8$, using $M_X=10^{16.2}$GeV and
$M_{\rm st}=10^{17.6}$GeV.
These combinations with $\Delta b'^k=4$ and 3 are classified into 13 and 15
types by values of $b'^i_3$ and $b'^i_2$ as shown in Table 7, where the values
of $b'^k_1$ are omitted.
The combinations with $\Delta b'^k=4$ derive $1\leq k_1 \leq 41/21$, including
 5/3, while the cases with $\Delta b'^k=3$ do not lead to 5/3.

The $Z_6$-II orbifold models allow 4586 combinations, which include a
combination with $\Delta b'^k=6$ as the largest value leading $T_k=6.5$.
Among the 4586 combinations, 632 ones with $\Delta b'\geq 4$ are classified
by values of $b'^i_3$ and $b'^i_2$ as shown in Tables 8-1 and 8-2, where
the values of $b'^k_1$ are omitted.
As shown in the tables, the minimal string model with $k_1=5/3$ cannot be
derived from $Z_6$-II orbifold.
However, if we permit the cases with $0 \leq \Delta b'^k < 4$, we can have
combinations of modular weights leading the measured couplings and $k_1=5/3$
in the case where $\Delta b'^k=2$ and 1.
The result coincides with ref.\cite{Ibanez}.

Note that any orbifold model lead to integer values of
$\Delta b'^k$, although the elements of the modular weights are fractional.
Types shown explicitly in Tables 6, 7, 8-1 and 8-2 are available for model
building.
In the above analysis, we have considered the case of the SUSY-breaking at
$M_Z$.
We can easily extend to other cases, e.g., the SUSY-breaking at 1TeV.

\vspace{0.8 cm}
\leftline{\large \bf 5. Hidden Sector}
\vspace{0.8 cm}

It is expected that the hidden sector plays a role in the SUSY-breaking.
The gaugino condensation in the hidden sector is one of the most attractive
proposals to break the SUSY.
In this scenario, the condensation around $10^{13}$GeV leads to soft
SUSY-breaking terms at the weak scale in the observable sector.
Therefore it is very important to investigate what kinds of hidden sectors are
consistent with the MSSM in the observable sector and to estimate the values
of the gauge coupling constants at $10^{13}$GeV in the hidden sectors.

For concreteness, we discuss the hidden sector of the minimal string
model of the $Z_6$-I with $\Delta b'^k=3$ shown in the second row of Table 6.
We take a SU($N)'$ group as the hidden sector.
If the hidden sector has no matter field with a non-trivial representation
under the SU($N)'$ group, we obtain $b'^i=-N$.
This hidden sector is not allowed, because $b'^j_3=-7/2$ ($j=1,2$).
Thus we need some hidden matter fields with non-trivial representations.
Especially for SU($N)'$ ($N>4$), we need matter fields with the modular
weights to increase $b'^j_{\rm hid}$, i.e., $n^i=(0,0,-1)$, which corresponds
to the untwisted sector associated with the third plane.
The other modular weights decrease the value of $b^j_{\rm hid}$.
Here we restrict ourselves to matter fields with $N$-dim fundamental
representations.
For $b'^j_{\rm hid}\geq -7/2$, we need the $M$ ($M\geq 2N-7$) matter fields
in the above untwisted sector.

For example, an SU(9)$'$ hidden gauge group must have at least eleven matter
fields with the 9-dim representation in the untwisted sector.
However such a larger number of the matter fields seems to be unrealistic.
Actually in the case with vanishing Wilson lines the untwisted matter
spectra are shown in ref.\cite{Z46} and those spectra do not include such a
large number of the matter fields with 9-dim representation.
If the Wilson lines do not change the gauge group, we can easily estimate
the number of the matter fields in the case of non-vanishing Wilson lines.
The presence of the Wilson lines restricts the matter fields.
Therefore the number of the matter fields without Wilson lines is the maximum
 among the general cases.
Further we have to consider the case that the Wilson lines break gauge groups.
Namely we can derive the SU(9)$'$ group by the Wilson lines from a large group.
Even including these cases, we cannot find eleven or more matter fields, but
at most only the three matter fields are allowed.
Similarly we cannot obtain nine (seven) or more matter fields with the
8-dim (7-dim) representation of the SU(8)$'$ (SU(7)$'$) group.
Thus the minimal string model from the $Z_6$-I with $\Delta b'^k=3$ cannot
have SU($N)'$ ($N>6$) hidden gauge groups.
On the other hand, we have the possibilty to obtain the required number of
the matter fields for the other SU($N)'$ hidden groups.
The first and fourth columns of Table 9 show the smallest numbers of the
matter fields in the hidden sectors satisfying the duality anomaly
cancellation condition with $b'^j_3=-7/2$.
In the column ($N,M$) shows the $M$ matter fields with the $N$-dim
representation under the SU($N)'$ hidden group.

Next we estimate the gauge coupling constants of the hidden groups allowed by
 the above arguments.
We eliminate $\delta^k_{\rm GS}$ and $T_k$ in eq.(3.10) for the hidden gauge
couplings using the SU(2) and SU(3) gauge couplings.
Then we obtain the following equation \cite{koba},
$$ \alpha_{\rm hid}^{-1}(\mu)=\alpha_X^{-1}+{1\over 4\pi}b_{\rm hid}\log
{M_X^2\over \mu^2}+{1\over 4\pi}\left( b_{\rm hid}-1+4
{b'^k_{\rm hid}-b'^k_2\over \Delta b'^k}\right)
\log {M_{\rm st}^2\over M_X^2}, \eqno(5.1) $$
where $b_{\rm hid}$ is an N=1 $\beta$-function coefficient for the hidden
group, i.e., $b_{\rm hid}=-3N+M$ for the $M$ matter fields in the SU($N)'$
group.
Note that $M_X$ means the unification scale of SU(3) and SU(2) in the
observable sector, and the gauge coupling of the hidden sector does not
always unify with the SU(3) and SU(2) gauge couplings at $M_X$.
We use $\alpha^{-1}_X=24.5$ at $M_X=10^{16.2}$GeV to estimate
$\alpha^{-1}_{\rm hid}$ at $10^{13.0}$GeV.
The results are listed in the second and fifth columns of Table 9, while the
third and sixth columns show blow-up scales $\Lambda$ (GeV) where
$\alpha^{-1}_{\rm hid}(\Lambda)=0$.

Similarly we can find the hidden sector satisfying the duality anomaly
cancellation condition with the minimal string model in the case of the
$Z_8$-I orbifold.
For simplicity we consider here some smaller numbers of the matter fields in
the hidden sector.
Tables 10-1, 10-2 and 10-3 show only allowed hidden sectors where the numbers
of the matter fields are less than 3, 6 and 11 for SU(2)$'$, SU(3)$'$ and
SU($N)'$ ($N>3$), respectively.
For the case with the three matter fields in SU(2)$'$, the gauge coupling
constant does not blow up at higher than 100GeV.
The larger hidden group requires the larger number of the matter fields.
The second column of Tables 10-1, 10-2 and 10-3 shows the least number of the
matter fields in a gauge group for each type of minimal string model with
$\Delta b'^k=4$ found in Table 7, and the number in the parentheses
corresponds to the case with $\Delta b'^k=3$.
For example, Type 3 with $\Delta b'^k=4$ is allowed to have the $M$
($M\geq 3$) matter fields with the 4-dim representations of the hidden
SU(4)$'$ group.
Similarly the least numbers of the $N$-dim hidden matter fields for Type 3 are
5, 7 and 9 for the SU(5)$'$, SU(6)$'$ and SU(7)$'$ hidden groups, respectively.
Type 3 does not allow the ten or less matter fields with the 8-dim
representation of the SU(8)$'$ group.
The hidden sectors are constrained much more severely than the case of
ref.\cite{koba},
 where is studied the duality anomaly cancellation condition
for the observable SU(3), SU(2) and hidden groups.
Note that it is difficult for some types to have the hidden sectors
such as considered here.
Especially Types 2, 5 and 9$\sim$ 15 with $\Delta b'^k=3$ cannot have hidden
sectors with $M$ ($M<11$) matter fields under SU($N)'$ ($N>3$).

In addition to $n^i=(0,0,-1)$, the $Z_8$-I and $Z_6$-II orbifold models have
the modular weights increasing the values of $b'^j_{\rm hid}$, which
correspond to the twisted sectors.
Therefore we do not have a constraint from the largest number of the untwisted
matter fields, which is used for the $Z_6$-I orbifold models.

Next we can estimate the gauge coupling constants of the above hidden groups.
The third and fourth columns of Tables 10-1, 10-2 and 10-3 show the gauge
couplings $\alpha^{-1}_{\rm hid}$ at $10^{13.0}$GeV and the blow-up scales
$\Lambda$ (GeV) in the case of $\Delta b'=4$, respectively.
The numbers in the parentheses correspond to the case of $\Delta b'=3$.
Note that the maximum values of $\alpha^{-1}_{\rm hid}(10^{13.0}$GeV) increase
by 1.1 as the number of the matter fields becomes larger by one.
Even in the case of the sixteen matter fields of the SU(8)$'$, gauge coupling
blows up at higher than $10^{13.4}$GeV.

Similarly we can find the hidden sectors and their gauge coupling constants
of the $Z_6$-II minimal string models.
Tables 11-1 and 11-2 show less than 4, 6 and 11 matter fields with the 2-dim,
3-dim and $N$-dim representation in the SU(2)$'$, SU(3)$'$ and SU($N)'$
($N\geq 4$), respectively.
Forbidden hidden sectors are omitted there.
The second, fourth and sixth columns show the gauge coupling constants at
$10^{13.0}$GeV, consistent with the minimal string models which have
$\Delta b'=6,5$ and 4, respectively.
Numbers in the third, fifth and seventh columns correspond to blow-up scales
$\Lambda$ (GeV) of the hidden gauge coupling constants.
We need the thirteen matter fields with the 7-dim representation for the gauge
coupling constants of SU(7)$'$ not so as to blow up at higher than
$10^{13.0}$GeV.
Even if we consider the case with the fourteen matter fields with the
fundamental representation, the gauge coupling constants of SU(8)$'$ and
SU(9)$'$ blow up at higher than $10^{13.7}$GeV and $10^{14.4}$GeV,
respectively.

\vspace{0.8 cm}
\leftline{\large \bf 6. Yukawa Coupling}
\vspace{0.8 cm}

In this section, we study Yukawa couplings as a further phenomenological
application of the above models.
The orbifold models have restrictive selection rules for the Yukawa couplings
\cite{Yukawa,KO2,KO}.
A point group selection rule requires that a product of point group elements
should be an identity.
Sectors allowed to couple are shown explicitly in ref.\cite{KO2,KO}.
Further the $Z_N$ invariance requires a product of the $Z_N$ phases from the
oscillated states to be zero.
Thus the Yukawa couplings of the oscillated states are much more restrictive.

First of all, we apply the above selection rules to the $Z_6$-I minimal
string model with $\Delta b'^k=3$.
Then we find that the selection rules do not allow any coupling.
Next we analyze the $Z_8$-I minimal sting models similarly.
Allowed couplings of the sectors are obtained as follows,
$$U_1U_2U_3,\quad T_1T_2T_5, \quad T_2T_2T_4, \quad U_2T_4T_4,
\eqno(6.1)$$
where $U_i$ is the untwisted sector associated with the $i$-th plane and
$T_\ell$ represents $\theta^\ell$-twisted sector.
The combination of the modular weights corresponding to Type 6 with
$\Delta b'^k=4$ allow only the Yukawa coupling for the top quark.
The other types with $\Delta b'^k=4$ do not permit any couplings.
The top Yukawa coupling is allowed by a combination of ($U_3,U_3,T_5$)
for $Q$, ($U_1,U_3,T_5$) for $U$,
($U_2,\tilde T_1(N_1=2/8),\tilde T_5(N_1=2/8),\tilde T_5(N_3=2/8),
\tilde T_5(N_3=2/8),$) for $L$ and $H$, and ($T_2,T_5,T_5$) for $E$, where
$\tilde T_\ell(N_i)$ indicates the $\theta^\ell$-twisted sector with the
oscillation number $N_i$ corresponding to the $i$-th plane.
Note that the hidden sectors of Type 6 are very constrained.

We can investigate the Yukawa couplings of the $Z_8$-I minimal string models
with $\Delta b'^k=3$ in the similar way.
Types 4,11,13 and 14 permit a Yukawa coupling for the top quark alone, and
Type 9 permit Yukawa couplings for the top and bottom quarks.
The others do not allow any coupling.
We remark that among types allowing the couplings only Type 4 appears
in Tables 10-1, 10-2 and 10-3, i.e., only Type 4 is possible to have a rich
structure in the hidden sector.

Similarly we study the Yukawa couplings of the $Z_6$-II minimal string
models, where we have the following couplings,
$$U_1U_2U_3, \quad T_1T_2T_3, \quad T_1T_1T_4, \quad T_2T_4U_3,
\quad T_3T_3U_2.
\eqno(6.2)$$
For $\Delta b'^k \geq 5$, Type 3, 4 and 8 allow only the top Yukawa coupling
and the other types permit no coupling.
For $\Delta b'^k =4$, most of types allow only the top Yukawa coupling but
Type 12 allows the top and bottom Yukawa couplings.
In Type 12 three generations of the quark doublets, $Q$ are assigned to
($U_{1-3},U_{1-3},T_3$), where $U_{1-3}$ denotes $U_1$ or $U_3$.
Further $U$, $D$ and $E$ belong to ($T_2,T_3,\tilde T_4(N_2=2/6)$),
($T_2,\tilde T_2(N_1=2/6), \tilde T_4(N_2=2/6)$) and ($U_2,U_2,T_2$),
 respectively.
Anomg $L$ and $H$, three matter fields belong to $T_4$ and the others belong
to $T_1$, and further three of the five have oscillation number $N_1=2/6$.

In the above models the Yukawa couplings are fairly constrained.
However, we remark that nonrenormalizable coupling could also lead
other Yukawa couplings.
These couplings are considered to be suppressed by at most $1/M_{\rm st}$.

\vspace{0.8 cm}
\leftline{\large \bf 7. Conclusion and Discussion}
\vspace{0.8 cm}

In this paper we have studied the possibility of obtaining the MSSM with
the measured values of the gauge coupling constants from
the $Z_N$ orbifold models.
We have used the duality anomaly cancellation condition and the
moduli dependent threshold corrections to the gauge coupling constants.
We have restricted ourselves to the case where the expectation value of the
moduli field is of order one.
Under the restriction, the $Z_8$-I and $Z_6$-II orbifold models are very
promising.
Allowed combinations of the modular weights are shown in Tables 6, 7, 8-1 and
8-2.
That is very useful for model building.
We have also found that the minimal string models on the $Z_6$-I, $Z_6$-II and
 $Z_8$-I orbifolds are possible to have the levels of
$k_1=29/21$, $1\leq k_1\leq 32/21$ and $1\leq k_1\leq 41/21$,
 respectively, for $\Delta b'^k\geq 3$.
Although the GUT prediction value $k_1=5/3$ is included only in $Z_8$-I, it is
 also included in the $Z_6$-II as discussed in ref.~\cite{Ibanez} if $\Delta
 b'^k=2$ or 1 is permitted.
Further we have the lower bounds for the values of $T_k$ as
$T_k\geq 11$, 6.5 and 8.8 in the minimal string models derived from
$Z_6$-I, $Z_6$-II and $Z_8$-I orbifold constructions, respectively.
Note that any orbifold model always derive integer values of
$\Delta b'^k$ in spite of fractional values of the modular weights.

In this paper we have assumed that the soft SUSY-breaking masses are universal
and of order $M_Z$.
However the string theories in general derive non-universal soft masses
\cite{Ibanez,non}.
Ref.\cite{KSY} shows that the non-universality of the soft masses is important
for analyzing the gauge coupling unification.
The non-universality often increases the unification scale $M_X$ and makes it
possible that smaller threshold corrections could explain the measured
values of the gauge coupling constants.
In this non-universal cases the smaller values of $\Delta b'^k$, e.g.
$\Delta b'^k=1$ or 2 could lead to $T$ of order one.
Therefore it is very intriguing to analyze similarly as the above
with including the non-universality of the soft SUSY-breaking masses.

We have also studied the hidden sectors of the minimal string models.
The structures of the hidden sectors are also strongly constrained by the
duality anomaly cancellation condition.
Actually, hidden sectors with smaller number of the matter fields are often
ruled out.
We have restricted ourselves to the case of the hidden sectors with the
SU($N)'$ gauge groups.
We can easily extend other hidden gauge groups like SO($2N)'$ and E$_N'$.

At last we have discussed the Yukawa couplings allowed in the above minimal
string models.
The condition of the Yukawa couplings can be used as a phenomenological
constraint for the minimal string models.
The constraint on the Yukawa couplings is discussed for the $Z_N \times Z_M$
orbifold models in ref.\cite{koba2}.

Although we have not investigated $Z_{12}$-I and $Z_{12}$-II orbifold models,
 the above procedure can be also applied to them.
One will be able also to investigate the supersymmetric standard models with
some extensions by extra matters \cite{extra} through the similar estimations,
although singlets like the right-handed neutrinos have been included
in the above discussions.
Further string models could be expected to have some extra U(1)'s in general.
Thus it is interesting to extend the above analyses including models with
extra U(1)'s.
Inclusion of the extra U(1)'s leads another constraint due to the duality
anomaly cancellation condition.

\vspace{0.8 cm}
\leftline{\large \bf Acknowledgement}
\vspace{0.8 cm}

The authors would like to thank D.~Suematsu and Y.~Yamagishi for useful
discussions.
The work of T.K. is supported in part by Soryuushi Shogakukai.


\newpage
\pagestyle{empty}
\noindent
\begin{center}
{\bf \large Table 1. modular weights in twisted sectors on $Z_4$ orbifold}
\vspace{10mm}

\renewcommand{\arraystretch}{1.5}
\begin{tabular}{|c||c|c|c|c|c|c|c|}
\hline
 Twisted  & $4n^i$ & \multicolumn{5}{c|}{Lower-bound of $k_1$} \\
\cline{3-7}
 sector   &    & $(3,2)_{1/6}$ & $(\bar 3,1)_{1/3}$
 & $(\bar 3,1)_{-2/3}$ & $(1,2)_{\pm 1/2}$ & $(1,1)_1$   \\ \hline \hline
$\theta$  & $(-3,-3,-2)$           & 4/15 & 16/51 &  64/51 & 4/7  & 16/11 \\
          & $(\underline{-7,-3},-2)$  & - & 16/15 & -  & 4/3  & - \\
 \hline
$\theta^2$& $(-2,-2,0)$            & 1/6  & 4/15  &  16/15 & 1/2  & 4/3  \\
 \hline
                                   \end{tabular}

\end{center}
\noindent

\begin{center}
{\bf \large Table 2. Modular weights in twisted sectors on $Z_6$-I orbifold}
\vspace{10mm}

\renewcommand{\arraystretch}{1.5}
\begin{tabular}{|c||c|c|c|c|c|c|c|c|}
\hline
 Twisted  & $6n^i$ & \multicolumn{5}{c|}{Lower-bound of $k_1$} &SU($N)'$ \\
\cline{3-7}
 sector   &    & $(3,2)_{1/6}$ & $(\bar 3,1)_{1/3}$
 & $(\bar 3,1)_{-2/3}$ & $(1,2)_{\pm 1/2}$ & $(1,1)_1$  & $\underline N$
\\ \hline \hline
$\theta$  & $(-5,-5,-2)$           & 1/6  & 4/15  &  16/15 & 1/2  & 4/3  & +
\\
          & $(\underline{-11,-5},-2)$  & - & 4/9  &  16/9  & 3/4  & 12/7 & +
\\
          & $(\underline{-17,-5},-2)$ & - &  4/3  &    -   & 3/2  &   -  &
$N\leq 6$ \\
          & $(-11,-11,-2)$            & - &  4/3  &    -   & 3/2  &   -  &
$N\leq 6$ \\
          & $(-5,-5,4)$               & - &  4/3  &    -   & 3/2  &   -  &
$N\leq 6$ \\
 \hline
$\theta^2$& $(-4,-4,-4)$           & 1/3  & 1/3   &  4/3   & 3/5  & 3/2  & +
\\  \hline
$\theta^3$& $(-3,-3,0)$            & 1/6  & 4/15  & 16/15  & 1/2  & 4/3  & +
\\  \hline
                                   \end{tabular}

\end{center}
\newpage
\begin{center}
\noindent

{\bf \large Table 3. Modular weights in twisted sectors on $Z_6$-II orbifold}
\vspace{10mm}

\renewcommand{\arraystretch}{1.5}
\begin{tabular}{|c||c|c|c|c|c|c|c|c|}
\hline
 Twisted  & $6n^i$ & \multicolumn{5}{c|}{Lower-bound of $k_1$} & SU($N)'$ \\
\cline{3-7}
 sector   &   & $(3,2)_{1/6}$ & $(\bar 3,1)_{1/3}$
 & $(\bar 3,1)_{-2/3}$ & $(1,2)_{\pm 1/2}$ & $(1,1)_1$ &$\underline N$ \\
\hline \hline
$\theta$  & $(-5,-4,-3)$ & 1/3  & 4/13  &  16/13  & 9/16 & 36/25 & + \\
          & $(-11,-4,-3)$ & -   & 4/7  &     -    & 9/10 & 36/19 & + \\
 \hline
$\theta^2$& $(-4,-2,0)$  & 1/7  & 1/5   &  4/5    & 9/19 & 9/7 & +  \\
          & $(-10,-2,0)$ &  -   & 1/2  &   2      & 9/7  &  -   & $N=2$ \\
          & $(-4,4,0)$   &  -   & 1/2  &   2      & 9/7  &  -   & $N=2$ \\
 \hline
$\theta^3$& $(-3,0,-3)$   & 1/6  & 4/15  & 16/15  & 1/2  & 4/3  & + \\
 \hline
$\theta^4$& $(-2,-4,0)$  & 1/7  & 1/5   &  4/5    & 9/19 & 9/7 & + \\
          & $(-2,-10,0)$ &  -   & 1/2  &   2      & 9/7  &  -  & $N=2$ \\
          & $(4,-4,0)$   &  -   & 1/2  &   2      & 9/7  &  -  & $N=2$ \\
 \hline
                                   \end{tabular}

\end{center}
\newpage
\noindent

\begin{center}
{\bf \large Table 4. Modular weights in twisted sectors on $Z_8$-I orbifold}
\vspace{10mm}

\renewcommand{\arraystretch}{1.5}
\begin{tabular}{|c||c|c|c|c|c|c|c|}
\hline
 Twisted  & $8n^i$ & \multicolumn{5}{c|}{Lower-bound of $k_1$} \\
\cline{3-7}
 sector   &  & $(3,2)_{1/6}$ & $(\bar 3,1)_{1/3}$
 & $(\bar 3,1)_{-2/3}$ & $(1,2)_{\pm 1/2}$ & $(1,1)_1$ \\ \hline \hline
$\theta$  & $(-7,-3,-6)$  & 16/87 & 64/231 & 256/231 & 16/31 & 64/47  \\
          & $(-15,-3,-6)$ & 16/15 & 64/159 & 256/159 & 16/23 & 64/39 \\
          & $(-23,-3,-6)$ &   -   & 64/87  &    -    & 16/15 &   -   \\
          & $(-7,-3,-14)$ &   -   & 64/87  &    -    & 16/15 &   -   \\
 \hline
$\theta^2$& $(-6,-6,-4)$  & 4/15  & 16/51  &  64/51  & 4/7 & 16/11  \\
          & $(\underline{-14,-6},-4)$ &
                               -    & 16/15  &       -  & 4/3  &    -  \\
 \hline
$\theta^4$& $(-4,-4,0)$            & 1/6  & 4/15  & 16/15  & 1/2  & 4/3 \\
 \hline
$\theta^5$& $(-3,-7,-6)$  & 16/87 & 64/231 & 256/231 & 16/31 & 64/47 \\
          & $(-3,-15,-6)$ & 16/15 & 64/159 & 256/159 & 16/23 & 64/39 \\
          & $(-3,-23,-6)$ &   -   & 64/87  &    -    & 16/15 &   -   \\
          & $(-3,-7,-14)$ &   -   & 64/87  &    -    & 16/15 &   -   \\
 \hline
                                   \end{tabular}

\end{center}
\newpage
\noindent

\begin{center}
{\bf \large Table 5. Modular weights in twisted sectors on $Z_8$-II orbifold}
\vspace{10mm}

\renewcommand{\arraystretch}{1.5}
\begin{tabular}{|c||c|c|c|c|c|c|c|}
\hline
 Twisted  & $8n^i$ & \multicolumn{5}{c|}{Lower-bound of $k_1$} \\
\cline{3-7}
 sector   &    & $(3,2)_{1/6}$ & $(\bar 3,1)_{1/3}$
 & $(\bar 3,1)_{-2/3}$ & $(1,2)_{\pm 1/2}$ & $(1,1)_1$ \\ \hline \hline
$\theta$  & $(-7,-5,-4)$  & 16/69 & 64/213 & 256/213 & 16/29 & 64/45 \\
          & $(-15,-5,-4)$ &   -   & 64/141 & 256/141 & 16/21 & 64/37 \\
          & $(-23,-5,-4)$ &   -   & 64/69  &    -    & 16/13 &   -   \\
 \hline
$\theta^2$& $(-6,-2,0)$  & 4/33  & 16/69  &  64/69  & 4/9 & 16/13  \\
          & $(-14,-2,0)$ &   -   & 16/33  &  64/33  & 4/5 & 16/9   \\
          & $(-6,6,0)$   &   -   & 16/33  &  64/33  & 4/5 & 16/9   \\
 \hline
$\theta^3$& $(-5,-7,-4)$  & 16/69 & 64/213 & 256/213 & 16/29 & 64/45 \\
          & $(-5,-15,-4)$ &   -   & 64/141 & 256/141 & 16/21 & 64/37 \\
          & $(-5,-23,-4)$ &   -   & 64/69  &    -    & 16/13 &   -   \\
 \hline
$\theta^4$& $(-4,-4,0)$            & 1/6  & 4/15  & 16/15  & 1/2  & 4/3  \\
 \hline
$\theta^6$& $(-2,-6,0)$  & 4/33  & 16/69  &  64/69  & 4/9 & 16/13  \\
          & $(-2,-14,0)$ &   -   & 16/33  &  64/33  & 4/5 & 16/9   \\
          & $(6,-6,0)$   &   -   & 16/33  &  64/33  & 4/5 & 16/9   \\
 \hline
                                   \end{tabular}

\end{center}

\newpage
\normalsize
\begin{center}
{\bf \large Table 6. Minimal String Model from $Z_6$-I orbifold}
\vspace{5mm}

\begin{tabular}{|c|c|c|c|c|c|c|}
\hline
 \#  & $\Delta b'^k$ ($T$) & $Q$  & $U$ & $D$ & $L$, $H$ & $E$ \\
\hline \hline
  1  &  3 (11)    & 3,3,4 & 4,4,6 & 9,9,9 & 6,7,7,8,8 & 3,3,4 \\
  2  &  2 (15)    & 1,2,3 & 5,6,6 & 9,9,9 & 3,7,7,8,8 & 3,4,4 \\
  3  &  2 (15)    & 3,5,5 & 5,6,6 & 9,9,9 & 3,7,7,8,8 & 3,4,4 \\
  4  &  1 (28)    & 1,2,3 & 4,4,4 & 4,9,9 & 6,7,7,8,8 & 3,3,4 \\
  5  &  1 (28)    & 3,5,5 & 4,4,4 & 4,9,9 & 6,7,7,8,8 & 3,3,4 \\
\hline
                                   \end{tabular}

\newpage
\normalsize
{\bf \large Table 7. Minimal String Model from $Z_8$-I orbifold }
\vspace{5mm}

\begin{tabular}{|c|c|c|c||c|c|c|c|}
\hline
\multicolumn{4}{|c||}{$\Delta b'^k=4$} &
\multicolumn{4}{|c|}{$\Delta b'^k=3$ }\\ \hline
 Type  & $k_1$  & $(8b'^1_3,8b'^2_3)$ & $8b'^3_3$
& Type  & $k_1$  & $(8b'^1_3,8b'^2_3)$ &$8b'^k_3$ \\  \hline \hline
  1  &   1     & (-12,-12) & -16 & 1 & 15/16 & (-28,-28) & -40  \\
  2  &   23/21 & (-22,-22) & -12 & 2 &  7/6  & (-28,-12) & -32  \\
  3  &   7/6   & (-36,-20) & -32 & 3 & 7/6   & (4,-44)   & -32 \\
  4  &   7/6   & (-4,-52)  & -32 & 4 & 19/15 & (-28,-28) & -40 \\
  5  &   53/39 & (-42,-42) & -28 & 5 & 4/3   & (-20,-20) & -16 \\
  6  &  29/21  & (-1,-29)  & -34 & 6 & 4/3   & (4,-44)   & -16 \\
  7  &   5/3   & (-55,-11) & -38 & 7 & 4/3   & (-4,-52)  & -48 \\
  8  &  5/3    & (-19,-47) & -38 & 8 & 53/39 & (-34,-34) & -28 \\
  9  &  5/3    & (-51,-15) & -38 & 9 & 29/21 & (-14,-14) & -28 \\
  10  & 5/3    & (-27,-39) & -38 & 10& 29/21 & (-23,-51) & -38 \\
  11 & 5/3     & (-59,-7)  & -38 & 11& 29/21 & (-22,-22) & -36 \\
  12 &  5/3    & (-23,-43) & -38 & 12& 17/12 & (-40,-8)  & -48 \\
  13 &  41/21  & (-23,5)  & -46  & 13& 17/33 & (-34,-34) & -12 \\
     &       &    &      &         14&  17/9 & (1,-35)   & -38 \\
     &       &    &      &         15& 41/21 & (-43,-15) & 2 \\
\hline
                                  \end{tabular}

\newpage
\normalsize
{\bf \large Table 8-1. Minimal String Model from $Z_6$-II orbifold
with $\Delta b'^k=6$ and 5}
\end{center}
Typle 1 has $\Delta b'^k=6$ and the other correspond to
$\Delta b'^k=5$ and $T=7.4$
\vspace{5mm}
\begin{center}

\begin{tabular}{|c|c|c|c|c|}
\hline
 Type  & $k_1$  & $6b'^1_3$ & $6b'^k_3$ & $18b'^k_1$
\\  \hline \hline
  1  &   7/6 & 8 & 4 & 248  \\ \hline
  2  &   7/6 & 8 & -2 & 176,200,212  \\
  3  &       & 8 & 4  & 194,230,266,302  \\
  4  &       & 16 & -4  & 178  \\ \hline
  5  & 11/9  & 9 & -3  & 135  \\ \hline
  6  & 26/21  & 14 & -8  & 80  \\ \hline
  7  & 19/15  & 5 & 1 & 77,113,149  \\ \hline
  8  & 31/24  & 16 & -4 & 64,100  \\ \hline
  9  & 67/51  & 17 & -11 & 5,29  \\ \hline
  10  & 4/3  & 6 & 0 & 36  \\
  11 &   & 12 & -6 & 12  \\ \hline
  12 & 41/30 & 20 & -14 & -82,-46  \\ \hline
  13 & 29/21 & 14 & -8 & -58  \\ \hline
                                   \end{tabular}

\newpage
\normalsize
{\bf \large Table 8-2. Minimal String Model from $Z_6$-II orbifold
with $\Delta b'^k=4$}
\vspace{5mm}

\begin{tabular}{|c|c|c|c||c|c|c|c|}
\hline
 Type  & $k_1$  & $6b'^1_3$ & $6b'^k_3$  & Type  & $k_1$  & $6b'^1_3$ &
$6b'^k_3$ \\  \hline \hline
  1  &   16/15 & 10 & -4   &   15 & 31/24 & 16 & -10   \\
  2  &   23/21 & 7 & -1   &   16 & 43/33 & 11 & -11,-5    \\
  3  &   37/33 & 11& -5   &   17 &     & 22 & -16 \\
  4  &   7/6   & 4 & -10,2   &  18 & 67/51 & 17 & -17,-11 \\
  5  &         & 8 & -8,-2,4 &   19 & 4/3 & 6 & -6,0 \\
  6  &         & 16 & -10,-4 &  20 &     & 12 & -12,-6 \\
  7  &  61/51  & 17 & -11 &  21 &     & 18 & -12 \\
  8  & 47/39   & 13 & -7  &  22 & 77/57  & 19 & -13 \\
  9  & 11/9    & 9 & -9,-3 &  23 & 53/39  & 13 & -7 \\
  10 & 26/21  & 14 & -14,-8 &  24 & 41/30  & 20 & -20,-14 \\
  11  & 71/57  & 19 & -13 &  25 & 29/21  & 7 & -19 \\
  12 & 19/15  & 5 & -17,-5,1 &   26 &        & 14 & -14,-8 \\
  13 &        & 10 & -5,-4  &   27 & 46/33  & 22 & -16 \\
  14 &        & 20 & -14  &   28 & 17/12  & 8 & -8,-2 \\
\hline
                                  \end{tabular}

\newpage
\normalsize
{\bf \large Table 9. Hidden gauge coupling in $Z_6$-I orbifold}
\vspace{5mm}

\begin{tabular}{|c|c|c||c|c|c|}
\hline
 N,M& $\alpha^{-1}_{\rm hid}$   & $\log_{10} \Lambda$   &
 N,M& $\alpha^{-1}_{\rm hid}$   & $\log_{10} \Lambda$ \\  \hline \hline
  2,2  &  13.7--14.3  &  5.5--5.2 &  5,3  &  3.05       &  12.4 \\
  2,3  &  14.2--14.9  &  4.4--4.0 &  5,4  &  3.56       &  12.3 \\
  3,2  &  8.66--9.97  &  10.1--9.8&  5,5  &  4.07       &  12.1 \\
  3,3  &  9.17--11.1  &  9.7--9.0 &  5,6  &  4.58--5.89  &  12.0--11.7 \\
  4,1  &  5.09       &  11.8 &     5,7  &  5.09--6.40  &  11.8--11.4 \\
  4,2  &  5.60       &  11.6 &     6,5  &  1.01       &  12.8 \\
  4,3  &  6.11       &  11.4 &     6,6  &  1.52       &  12.8 \\
  4,4  &  6.62--7.93  &  11.2--10.8 & 6,7 &  2.03       &  12.6 \\
  4,5  &  7.13--8.44  &  11.0--10.6 & 6,8 &  2.54--3.85  &  12.5--12.3 \\
       &             &            & 6,9 &  3.05--4.36  &  12.4--12.1 \\ \hline
                                   \end{tabular}

\newpage
\normalsize
{\bf \large Table 10-1. Hidden gauge coupling in $Z_8$-I orbifold}
\vspace{5mm}

\begin{tabular}{|c|c|c|c|}
\hline
$N,M$ & Type & $\alpha^{-1}_{\rm hid}$ & log$_{10}\Lambda$ \\ \hline \hline
2,1 & 1 & 16.5 & 4.8 \\
2,2 & 2,3,5,9,12 (8) & 17.6--19.1 (19.9) & 2.6--3.4 (2.1) \\ \hline
3,1 & 12 & 12.5 & 9.0 \\
3,2 & 2,5,8,10 (1,4,8) & 11.6--13.6 (12.1--14.2) & 8.4--9.1 (8.1--8.9) \\
3,3 & 1,3,9            & 11.2--14.7 (12.7--15.4) & 7.7--8.9 (7.4--8.3) \\
3,4 & 4,6,7,11 (7)     & 13.3--15.7 (12.9--16.6) & 6.9--8.2 (6.5--8.0) \\
3,5 &          (6)     & 12.4--16.8 (13.8--17.7) & 5.9--7.8 (5.6--7.2) \\
 \hline
4,1 & (8)         &          (8.0)       & (11.1)  \\
4,2 & 5,10        & 7.1--8.1  (7.2--9.2)  & 11.0--11.3 (10.7--11.2)\\
4,3 & 3,12        & 8.2--9.1  (8.4--10.4) & 10.6--10.9 (10.3--10.8)\\
4,4 & 2,8 (1,4)   & 7.3--10.2 (8.8--11.5) & 10.2--11.0 (9.9--10.6)\\
4,5 & 1,9         & 6.9--11.3 (9.4--12.7) & 9.8--11.3  (9.4--10.3)\\
4,6 & 7           & 7.9--12.4 (8.6--13.9) & 9.2--10.7  (8.8--10.4)\\
4,7 & 4,11 (7)    & 8.0--13.5 (7.8--15.1) & 8.7--10.4  (8.2--10.5)\\
4,8 &             & 8.6--14.6 (9.0--16.2) & 8.0--10.1  (7.5--10.0)\\
4,9 & (3,6,8)     & 8.2--15.7 (7.5--18.0) & 7.3--10.1  (6.5--10.3)\\
4,10 & 6,13       & 7.8--16.7 (10.0--18.7)& 6.5--10.0  (5.7--9.1)\\
\hline
                                   \end{tabular}

\newpage
\normalsize
{\bf \large Table 10-2. Hidden gauge coupling in $Z_8$-I orbifold}
\vspace{5mm}

\begin{tabular}{|c|c|c|c|}
\hline
$N,M$ & Type & $\alpha^{-1}_{\rm hid}$ & log$_{10}\Lambda$ \\ \hline \hline
5,1 & 5         & 1.0               & 12.8 \\
5,2 &           & 0.6--2.0           & 12.6--12.9\\
5,3 &  (1,4,8)  & 1.6--3.1 (3.4--3.6) & 12.4--12.7 (12.3)\\
5,4 & 10        & 2.7--4.2 (2.7--4.5) & 12.1--12.4 (12.1--12.5)\\
5,5 & 3,12      & 2.8--5.3 (3.7--5.7) & 11.8--12.4 (11.8--12.2)\\
5,6 & 2,8       & 2.4--6.4 (4.9--9.0) & 11.6--12.5 (11.0--11.9)\\
5,7 & 1,9       & 2.0--7.5 (4.7--9.6) & 11.2--12.5 (10.7--11.9)\\
5,8 & 7         & 3.1--8.6 (3.9--10.6)& 10.9--12.2 (10.3--12.0)\\
5,9 & 4,11 (7)  & 3.7--9.6 (3.1--11.7)& 10.5--12.0 (10.0--12.2)\\
5,10 &          & 4.3--10.7(4.3--13.6)& 10.1--11.8 (9.3--11.8)\\ \hline
6,3 & 5       & ---                & 13.6\\
6,4 &         & ---                & 13.4--13.7\\
6,5 & (1,4,8) & --- (0.0--1.1)      & 13.2--13.5 (12.8--13.2)\\
6,6 & 10      & --- (0.0--1.7)      & 13.0--13.3 (12.6--13.2)\\
6,7 & 3,12    & 0.0--1.0 (0.0--2.4) & 12.8--13.3 (12.6--13.2)\\
6,8 & 2,8     & 0.0--2.0 (0.2--6.5) & 12.6--13.4 (11.7--13.0)\\
6,9 & 1,9     & 0.0--3.1 (0.1--7.1) & 12.4--13.5 (11.7--13.0)\\
6,10 & 7      & 0.0--4.2 (0.0--8.2) & 12.1--13.3 (11.3--13.2)\\
\hline
                                   \end{tabular}

\newpage
\normalsize
{\bf \large Table 10-3. Hidden gauge coupling in $Z_8$-I orbifold}
\vspace{5mm}

\begin{tabular}{|c|c|c|c|}
\hline
$N,M$ & Type & $\alpha^{-1}_{\rm hid}$ & log$_{10}\Lambda$ \\ \hline \hline
7,5 & 5 & --- & 14.1\\
7,6 &  & --- & 14.0--14.2\\
7,7 & (1,4,8) & --- (---) & 13.9--14.1 (13.2--13.9)\\
7,8 & 10      & --- (---) & 13.7--14.0 (13.1--14.1)\\
7,9 & 3,12    & --- (---) & 13.6--14.0 (13.0--13.9)\\
7,10 & 2,8    & 0.0--4.2 (0.0--4.0) & 13.4--14.2 (12.3--13.8)\\ \hline
8,7 & 5 & --- & 14.6\\
8,8 &  & --- & 14.5--14.7\\
8,9 & (1,4,8) & --- (---) & 14.4--14.6 (13.5--14.5) \\
8,10 & 10     & --- (---) & 14.3--14.5 (13.5--14.7) \\
\hline
                                   \end{tabular}

\newpage
\normalsize
{\bf \large Table 11-1. Hidden gauge coupling in $Z_6$-II orbifold}
\vspace{5mm}

\begin{tabular}{|c|c|c|c|c|c|c|}
\hline
$N,M$ & \multicolumn{2}{|c|}{$\Delta b'=6$ }&
\multicolumn{2}{|c|}{$\Delta b'=5$ }&
\multicolumn{2}{|c|}{$\Delta b'=4$ }\\ \cline{2-7}
   & $\alpha^{-1}_{\rm hid}$ & log$_{10}\Lambda$
& $\alpha^{-1}_{\rm hid}$ & log$_{10}\Lambda$
& $\alpha^{-1}_{\rm hid}$ & log$_{10}\Lambda$ \\ \hline \hline
2,3 & 17.3 & 2.5            & 17.1--17.5 & 2.4--2.7 & 16.7--18.7 & 1.7--2.9\\
3,4 & ---  & ---            & 12.3 & 8.2            & 11.3--12.8 & 8.0--8.6\\
3,5 & 13.0--13.3 & 7.4--7.6 & 12.5--14.1 & 7.1--7.7 & 11.9--14.9 & 6.8--8.0\\
4,4 & ---    & ---          & ---  & ---            & 5.8--6.8 & 11.2--11.4\\
4,5 & ---    & ---          & 7.1--7.5 & 10.9--11.0 & 6.4--8.3 & 10.6--11.1\\
4,6 & 8.3--8.9 & 10.3--10.5 & 7.7--8.9 & 10.3--10.7 & 7.0--9.4 & 10.1--10.9\\
4,7 & 9.3--9.9 & 8.9--9.8   & 8.4--10.8 & 9.6--10.3 & 7.5--11.5 & 9.3--10.6\\
4,8 & 9.3--11.9 & 8.9--9.8  & 8.6--12.2 & 8.8--10.1 & 7.6--12.6 & 8.7--10.4\\
4,9 & 10.3--12.9 & 8.3--9.3 & 9.3--13.2 & 8.2--9.6  & 8.2--15.2 & 7.5--10.0\\
4,10 & 10.6--14.3 & 7.4--8.7& 9.9--15.1 & 7.1--9.1  & 8.3--16.2 & 6.7--9.8 \\
5,5 & --- & ---             & 1.7       & 12.6      & 0.8--2.3 & 12.5--12.8\\
5,6 & 2.6 & 12.3            & 2.3--3.1 & 12.3--12.5 & 1.4--3.4 & 12.2--12.7\\
5,7 & 3.9 & 12.1            & 3.0--4.2 & 12.0--12.3 & 2.0--5.0 & 11.8--12.5\\
5,8 & 3.9--5.6 & 11.6--12.0 & 3.2--5.6 & 11.6--12.2 & 2.1--6.1 & 11.5--12.5\\
5,9 & 4.9--6.6 & 11.3--11.7 & 3.8--7.4 & 11.1--12.0 & 2.7--8.2 & 10.9--12.3\\
5,10 & 5.3--8.6 & 10.7--11.6& 4.5--8.8 & 10.6--11.8 & 2.8--9.2 & 10.5--12.2\\
\hline
                                   \end{tabular}

\newpage
\normalsize
{\bf \large Table 11-2. Hidden gauge coupling in $Z_6$-II orbifold}
\vspace{5mm}

\begin{tabular}{|c|c|c|c|c|c|c|}
\hline
$N,M$ & \multicolumn{2}{|c|}{$\Delta b'=6$ }&
\multicolumn{2}{|c|}{$\Delta b'=5$ }&
\multicolumn{2}{|c|}{$\Delta b'=4$ }\\ \cline{2-7}
   & $\alpha^{-1}_{\rm hid}$ & log$_{10}\Lambda$
& $\alpha^{-1}_{\rm hid}$ & log$_{10}\Lambda$
& $\alpha^{-1}_{\rm hid}$ & log$_{10}\Lambda$ \\ \hline \hline
6,6 & ---  & ---            & --- & 13.6            & --- & 13.7\\
6,7 & ---  & ---            & --- & 13.3--13.5      & --- & 13.2--13.7\\
6,8 & --- & 13.1--13.3      & --- & 13.1--13.4      & 0.0--0.1 & 13.0--13.7\\
6,9 & 0.0--0.6 & 12.9--13.1 & 0.0--0.8 & 12.8--13.3 & 0.0--1.6 & 12.7--13.6\\
6,10 & 0.0--2.2 & 12.5--13.0& 0.0--2.2 & 12.5--13.2 & 0.0--2.7 & 12.4--13.6\\
7,7 & --- & ---        & --- & ---        & --- & 14.2--14.4\\
7,8 & --- & 14.0--14.1 & --- & 14.0--14.2 & --- & 14.0--14.4\\
7,9 & --- & 13.9--14.0 & --- & 13.8--14.2 & --- & 13.7--14.4\\
7,10& --- & 13.7--13.9 & --- & 13.6--14.1 & --- & 13.6--14.4\\
8,8 & --- & 14.7       & --- & 14.7--14.8 & --- & 14.8--15.0\\
8,9 & --- & 14.6--14.7 & --- & 14.6--14.7 & --- & 14.5--14.9\\
8,10& --- & 14.4--14.6 & --- & 14.4--14.7 & --- & 14.4--15.0\\
9,9 & --- & ---        & --- & 15.1--15.2 & --- & 15.2--15.4\\
9,10& --- & 14.9--15.0 & --- & 15.0--15.1 & --- & 15.1--15.4\\
\hline
                                   \end{tabular}
\end{center}

\end{document}